# A decatungstate-based Ionic liquid exhibiting very low dielectric constant suitable for acting as solvent and catalyst for oxidation of organic substrates

Yohan Martinetto,[a,b] Bruce Pégot,*[a] Catherine Roch,[a] Mohamed Haouas,[a] Betty Cottyn-Boitte,[b] Franck Camerel,[c] Jelena Jeftic,[c] Denis Morineau,[d] Emmanuel Magnier,[a] and Sébastien Floquet.*[a]

**In this contribution, a new POM-based ionic liquid, namely ($P_{6,6,6,14}$)$_4$[$W_{10}O_{32}$], was fully characterized. Its viscosity and its very low dielectric constant make this hybrid ionic liquid suitable to be used as solvent for organic transformations. As proof of concept, this unique ionic liquid combining solvent and catalyst properties was tested for catalytic oxidation of various alcohols and alkenes in presence of $H_2O_2$**

Polyoxometalate (POM) compounds constitute a wide family of anionic inorganic compounds which can be finely tuned at the molecular level. The choice of the constituents, including the counter cations, is determinant to design the final architectures. Therefore, owing to their structural and compositional versatility, they possess a vast range of properties and potential applications, notably in catalysis.[1-4] In this domain, POMs have been widely used for the oxidation of various organic substrates, very often in the presence of water solutions of hydrogen peroxide.[5,6] They can be used in heterogenous or homogeneous catalytic processes. However, homogeneous catalytic processes display serious shortcomings of catalyst recovery or reaction mixture separation. Besides, due to detrimental effects in the environment, research for alternative reaction media instead of conventional organic solvents has rapidly grown. For these purposes, a particular attention has been given to reusability of catalysts and use of ionic liquids as solvents in the reaction systems.

The coupling of POMs with some organic cations allows the design of hybrid compounds at the boundary between molecular chemistry and chemistry of materials.[7] Among these hybrid materials, most of them are solid salts or materials with a high or even indeterminate melting points and cannot be formally called ionic liquids, or more precisely POM-based ionic liquids, abbreviated POM-ILs.[8] Interestingly, among the hundreds of hybrids POMs reported in the literature, only a few dozen of them can be considered as real POM-ILs, the others has been suggested to be labelled IL-POMs[3] Besides, a close look reveals that only very few members of this restricted POM-ILs family are liquid at room temperature and their physical properties such as viscosity and dielectric constant are very rarely investigated.[3]

Considering the potential catalytic properties of the POMs and the solvent properties of Ionic liquids, the use of recyclable POM-ILs could resolve both issues mentioned above and we are convinced that such compounds are a promising answer to the problematics encountered in homogeneous catalysis. However, a better understanding of the physical properties of POM-IL is needed for this purpose. While the use of IL-POM dispersed on surfaces or hybrid POMs solubilized in ILs are reported, the use of liquid POM-ILs is still surprisingly rare in catalytic reactions[3,9] and to the best of our knowledge, only one example reports the use of a room temperature POM-IL for the oxidation of alcohols and alkenes into carboxylic acids.[10]

To bring new insights for the knowledge of physical properties of POM-ILs and evaluate the interest of such material for homogeneous catalysis, the purpose of this work is to synthetize and to fully characterize a new POM-IL, liquid at room temperature and exhibiting the required properties to act both as a good solvent and as a catalyst for oxidizing organic molecules.

To design the POM-IL, the choice of phosphonium counter cations is of great interest. Indeed, such type of cation yields the lowest melting points when combined to a POM, even if it is not the most common cations for ionic liquids.[3,11]

In this study, the decatungstate POM [$W_{10}O_{32}$]$^{4-}$, known to possess catalytic and photocatalytic properties,[12-14] was associated with four trihexyltetradecylphosphonium cations as described in the ESI.

The material was isolated as viscous colorless oil for which the FT-IR, TGA, EDX, ESI-MS, Ion chromatography and NMR experiments agree well with the expected formula ($P_{6,6,6,14}$)$_4$[$W_{10}O_{32}$].

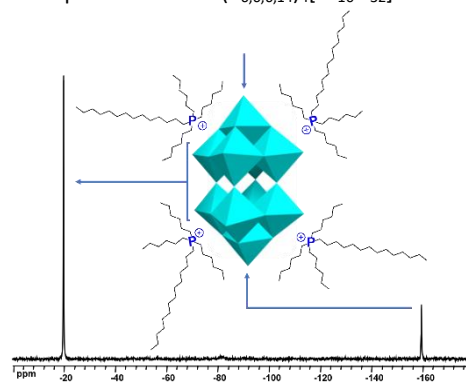

**Figure 1:** $^{183}$W NMR (16.7 MHz) spectrum of ($P_{6,6,6,14}$)$_4$[$W_{10}O_{32}$] in CD$_3$CN. Inset: Representation of the POM-IL ($P_{6,6,6,14}$)$_4$[$W_{10}O_{32}$]. Arrows corresponds to the assignments of the $^{183}$W NMR spectrum.

[a.] Institut Lavoisier de Versailles, UMR 8180, Université de Versailles St-Quentin en Yvelines, CNRS, Université Paris-Saclay, 78035 Versailles, France.
[b.] Institut Jean-Pierre Bourgin, INRAE, Agro Paris Tech, Université Paris Saclay, 78000 Versailles, France.
[c.] Univ Rennes, CNRS, ISCR (Institut des Sciences Chimiques de Rennes) - UMR 6226, 35000 Rennes, France.
[d.] Institut de Physique de Rennes, UMR 6251, Université de Rennes 1, 35042 Rennes, France.

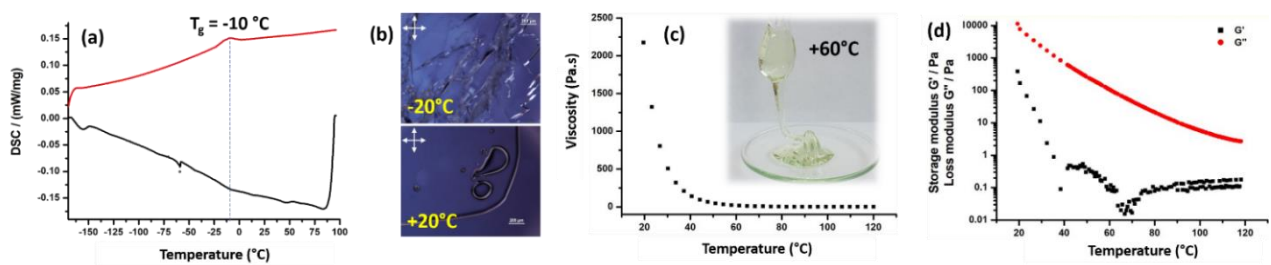

**Figure 2 :** (a) DSC curves of the (P$_{6,6,6,14}$)$_4$[W$_{10}$O$_{32}$] measured between 100 °C and -170 °C (Top: 2$^{nd}$ heating curve, bottom: 1$^{st}$ cooling curve, scan rate: 10 °C.min$^{-1}$; *:artefact due to the N$_2$ pumping); (b) Polarized optical microscopy images showing the edge of a drop of the compound between glass slides at -20°C and at +20°C after removal of the top cover slide; (c) Viscosity as a function of the temperature between 20 °C and 120 °C ($\dot{\gamma}$ = 10 s$^{-1}$); c) Temperature dependence of $G'$ and $G''$ (20% strain, f = 1Hz).

In particular, $^{183}$W NMR (Fig. 1 and Fig S5, SI) corresponds completely to the POM [W$_{10}$O$_{32}$]$^{4-}$,[14] while the $^1$H NMR spectrum (Fig. S4, ESI) shows that the signals of the methylenic groups of the counter-cations are significantly shifted because of the POM association, suggesting a strong interaction between the two components.

To demonstrate the true ionic liquid nature of this compound and to use it as a fluid solvent capable of solubilizing organic substrates, rheological and dielectric constant measurements were performed. Thermogravimetric analyses (Fig S3, ESI) show that the compound is stable up to 200 °C without any degradation and a clear glass transition centered around $T_g$ = -10 °C is observed on the DSC heating curves (Fig. 2a). Polarized Optical Microscopy observations show that, below $T_g$, the material is a brittle solid in an isotropic glassy state, while above $T_g$, the compound is viscous, isotropic and malleable material that spreads on a surface (Fig. 2b).[15]

To unambiguously confirm the liquid nature of this compound above room temperature, temperature-dependent rheological measurements were performed. The Figure S6 in ESI represents the evolution of the shear stress ($\tau$) as a function of the shear rate ($\dot{\gamma}$) at various temperatures. From the observed straight-line flow curves, it appears that the compound is a Newtonian fluid ($\tau = \eta \times \dot{\gamma}$, where $\eta$ is the viscosity) and that the viscosity (*i.e.* the slope of the curves) gradually decreases with the increase of the temperature. This was confirmed by the direct measurement of the viscosity as a function of the temperature (Fig. 2c). The viscosity measured around 2200 Pa.s at room temperature drastically and rapidly decreases when temperature increases. The viscosity is below 100 Pa.s above 40 °C and reaches 0.3 Pa.s at 120 °C. For comparison, the viscosity of ricin oil is 0.98 Pa.s at room temperature. The loss modulus ($G''$) and the storage modulus ($G'$) have also been measured as a function of the temperature between 20 and 120 °C (Fig. 2d). Both modulus values decrease with the temperature increase. $G''$ is at least two orders of magnitude larger than $G'$ over the whole temperature range explored, meaning that this compound is always in a quasi-liquid state. Generally, $G'$ concerns elastic properties of the material, while $G''$ represents the viscous nature of the compound, meaning that this compound exhibits a rather viscous behavior. It can also be noticed that the signal associated to the storage modulus ($G'$) is very weak and noisy, as expected for a liquid with a mainly viscous character. These results demonstrate that this compound is a highly viscous room-temperature ionic liquid (RTIL) which becomes more and more fluid as the temperature increases. This is fully compatible with fluid behavior needed for catalysis experiments.

To go further, we investigated the dielectric constant of this POM-LI, a property which is very rarely studied for POM-LIs. The evaluation of the solvation properties of the solvent is not unique and could be refined depending on the nature of the particular applications targeted.[16] Indeed, for ionic liquids, it was pointed out that the determination of polarity scales was influenced by the specificities of probe-solvent local interactions, thus challenging the relationship between polarity and static dielectric constant, which has been empirically established for conventional molecular solvents.[17] In our case, the complex dielectric function $\varepsilon^*(f) = \varepsilon'(f) - i\varepsilon''(f)$ was investigated from 10$^{-1}$ Hz to 10$^6$ Hz where f denotes the frequency, $\varepsilon'$ and $\varepsilon''$ the real and loss part of the complex dielectric function. The measurements were acquired on cooling from T = 70 °C to T = -150 °C with temperature steps of 5 °C. Depending on the temperature, different contributions were observed in the instrumental frequency window, as shown in Figure 3. At high temperature, e.g. T = 70 °C (Figure 3a), the intensity of the dielectric loss was dominated by a power law term (-1/f), which is commonly attributed to ionic conductivity.[18] On cooling, the conductivity contribution gradually decayed and two peaks emerged. They were attributed to dielectric relaxation modes, and denoted mode I and mode II, respectively (Figures 3b and 3c). The comprehensive depiction of the evolution of the dielectric loss for a wide range of temperatures is provided in Supplementary Information (Fig. S7-9). The dielectric measurements were analysed quantitatively by fitting the complex dielectric function by a model including two Havriliak and Negami functions (HN-model) and a conductivity term[19] according to eq. 1, where $\varepsilon_\infty$ is the sample permittivity in the limit of high frequency, $\Delta\varepsilon_{1,2}$ and $\tau_{HN1,2}$ are the dielectric strength and the relaxation time of the mode I (resp. II). $\sigma_0$ stands for the DC conductivity of the sample, where the exponent $s$=1 applies for Ohmic mechanisms.

$$\varepsilon^*(\omega) = \varepsilon_\infty + \sum_{k=1,2} \frac{\Delta\varepsilon_k}{(1+(i\omega\tau_{HNk})^{\beta_k})^{\gamma_k}} - i\frac{\sigma_0}{\omega^s \varepsilon_0} \quad (1)$$

According to the formalism of the HN-model, the exponent parameters $\beta$ and $\gamma$ are introduced to account for the symmetric and the asymmetric broadening of the complex dielectric function with respect to the Debye one, which is recovered when $\beta = \gamma = 1$.

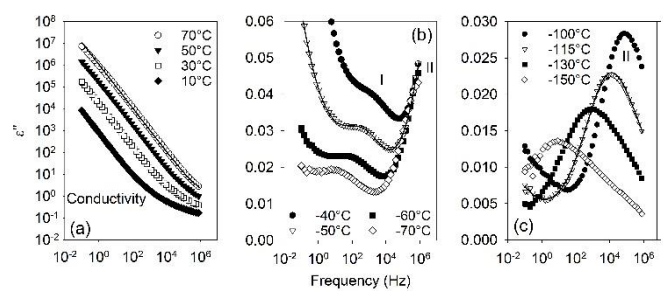

**Figure 3.** Dielectric loss as a function of the frequency at three different temperatures (a) 70°C, (b) -50°C and (c) -115°C. Solid lines denote fits by the Havriliak-Negami equations to the corresponding data (see text for details).

Although the application of these fractional parameters significantly improved the quality of the fits, their exact values were ambiguous because of the small value of the dielectric strength and the numerical degeneracy due to the coupling between $\beta$, $\gamma$ and the relaxation time $\tau_{HN}$. Therefore, we computed the value of the relaxation time $\tau_{max}$ corresponding to the maximum frequency of the relaxation peak in order to overcome this limitation, according to relation (2):

$$\tau_{max} = \tau_{HN} \left[\sin\left(\frac{\pi\beta}{2+2\gamma}\right)\right]^{-1/\beta} \left[\sin\left(\frac{\pi\beta\gamma}{2+2\gamma}\right)\right]^{1/\beta} \quad (2)$$

The resulting fits are shown as solid lines in Figure 3 for a selection of three temperatures. Equally good fits were obtained for all the studied temperatures. The real part of the dielectric permittivity was shown to be weakly temperature dependent, and estimated from a Cole-Cole plot (not-shown) to be $\varepsilon_\infty$ = 2.9 and $\varepsilon_s$ = 3.2 in the limit of high and low frequency, respectively, which agrees with viscosity measurements (see Fig. S9, ESI). This very low value is surprising if we consider the electrostatic nature of the assembly of the POM with the four phosphonium cations and confirms the strong interaction between these two components according to NMR data. These value compares well with organic solvents such as dioxane or diethyl ether ($\varepsilon$ = 2.25 and 4.33 respectively) and thus appear as good solvent for organic substrate despite its ionic nature.

This POM-IL exhibits the properties required to be used as a fluid solvent of organic transformation, *i.e.* a very low viscosity over 40°C, a dielectric permittivity around 3 and a thermal stability up to 200°C. The second part of this work was thus focused on the investigation of the catalytic behaviour of such a system. As a proof of concept, the oxidation reaction of alcohols and alkenes was studied in a preliminary study, to evaluate the efficiency of this new POM-IL as both solvent and catalyst. The oxidation of 4-fluorobenzyl alcohol, performed in a two-phase system (POM-IL phase+organic substrate / aqueous phase containing $H_2O_2$), was firstly optimized according to procedures described in ESI (Tables S1, ESI). As a first indication of the efficiency of our system, the alcohol was converted entirely into the corresponding carboxylic acid in several hours (93% isolated yield) as the sole compound. The aldehyde, which is certainly one of the intermediates, could be observed in minor proportion by reducing reaction time. Control reactions, without $H_2O_2$ or POM-IL, confirm the importance of the catalyst and its co-oxidant. Encouraged by this positive result, more challenging compounds were tested in the previous condition. The results are gathered in Table 1.

The process works efficiently on primary benzylic alcohols (entry 2) and also on secondary alcohols (entry 3). Interestingly, with cyclohexanol, the adipic acid, product of over-oxidation and C-C cleavage, was isolated as the major product without formation of the corresponding ketone (entry 4). This is in stark contrast with Guo results, who evidenced the oxidation of cyclohexanol to cyclohexanone at 90-100 °C for 2 hours with 89.1% yield by adding hexadecyltrimethylammonium decatungstate to the alcohol in presence of aqueous $H_2O_2$.[12]

This reinforces the idea that the oxidative cleavage of alkene is possible with such ionic liquid system. The first tests on cyclohexene under the previous conditions gave only traces of diacid. The heating mode was modified with the use of microwaves as activation which has shown some efficiency in that kind of catalysis.[20] Gratifyingly, in 2 hours with only 5 equivalents of $H_2O_2$, encouraging yields between 15 and 37% of the isolated main product could be obtained (entries 5 to 8). This microwave activation allowed the improvement of both the efficiency and the selectivity of the transformation with a concomitant drastic diminution of the quantity of hydrogen peroxide, reduced by a factor of ten. The moderate yields could be explained by the several intermediates formed during the oxidation processes (epoxide, diol, cyclic α-hydroxy ketone, α-hydroxy lactones).[6] Further optimizations of the reaction conditions are currently under development in our laboratory to push forward oxidation and get complete conversion into acids.

**Table 1.** scope of oxidation with $(P_{6,6,6,14})_4[W_{10}O_{32}]$ as catalyst[a]

| Entry | Substrate | Product | Yield (%)[b] |
|---|---|---|---|
| 1 | F-C6H4-CH2OH | F-C6H4-COOH | 93 |
| 2 | pyridyl-CH2OH | pyridyl-COOH | 89 |
| 3 | Ph2CHOH | Ph2C=O | 62 |
| 4 | cyclohexanol | adipic acid (HOOC-(CH2)4-COOH) | 15 |
| 5[c] | cyclohexene | adipic acid | 15 |
| 6[c] | 4-fluorostyrene | 4-fluorobenzoic acid | 37 |
| 7[c] | indene | 2-(carboxymethyl)benzoic acid | 28 |
| 8[c] | cinnamyl alcohol | benzoic acid derivative | 26 |

[a]Reaction conditions: alcohol (1 eq.), $H_2O_2$ 30 wt% (50 eq.), $(P_{6,6,6,14})_4[W_{10}O_{32}]$ (0.1 eq.), 90°C, 16h. [b]Isolated yield after catalyst/product separation by a reusable steric polymer column. [c]Reaction conditions: alcohol (1 eq.), $H_2O_2$ 30 wt% (5 eq.), $(P_{6,6,6,14})_4[W_{10}O_{32}]$ (0.1 eq.), MW irradiation, 75 °C, 90 min.

The traditional methods of extraction with organic solvents commonly used with ionic liquids, or even separation by column chromatography on silica gel, proved to be ineffective to separate the products from the POM-IL phase as well as to recycle the catalyst. Therefore, we demonstrated that the catalyst can be successfully separated from the mixture of organic products at the end of the reaction thanks to a reusable steric exclusion column. After elution of the reaction medium on this column, the POM-IL, non-retained on the solid phase thanks to its size, was firstly recovered and directly reused. The oxidation products are isolated secondly from the column and separated by classical chromatography on silica to get yields. The POM-IL phase was reused over several cycles as shown in Table 2. Over the first three cycles no loss of activity was observed:

the isolated yield was around 90%. Nevertheless, after 4 cycles, a slight loss of yield was observed. This phenomenon was partly explained by the formation of a less reactive species Lindqvist-type POM [$W_6O_{19}$]$^{2-}$ as the number of cycles increased (see Figure S9, $^{183}$W NMR, in ESI). Accordingly to the literature, the active species is likely a peroxotungstate species.[6] The formation of such intermediate active species involves the opening of the decatungstate POM moiety during the catalytic process, which can either lead to the initial product or to the Lindqvist product.

**Table 2**. Recycling of catalyst ($P_{6,6,6,14}$)$_4$[$W_{10}O_{32}$]. Isolated yields obtained for runs 1-5. Reaction conditions: 4-fluorobenzyl alcohol (1 eq.), $H_2O_2$ 30 *wt*% (50 eq.), ($P_{6,6,6,14}$)$_4$[$W_{10}O_{32}$] (0.1 eq.), 90 °C, 16 h.

| Entry | Run 1 | Run 2 | Run 3 | Run 4 | Run 5 |
|---|---|---|---|---|---|
| 1 | 93% | 92% | 86% | 73% | 68% |

In summary, in this contribution, we synthetized a new compound of formula ($P_{6,6,6,14}$)$_4$[$W_{10}O_{32}$]. The latter was unambiguously characterized as a newtonian ionic liquid with a glass transition at -10°C and rheological measurements allowed to evidence a good fluidity when temperature increased and a very low dielectric constant comparable with those obtained for apolar organic solvents. Used as solvent for organic molecules, the addition of an aqueous phase containing $H_2O_2$ evidenced the potentialities of such system for oxidation of alcohols and alkenes, in particular with C-C and C=C bond activations. Besides, we developped also an elegant way for recycling the POM-IL phase by using steric exclusion gel. This system appears to be a promising powerful catalyst and prompts us now to investigate its activity towards valorizable biopolymers considered so far as wastes by industry.

## Conflicts of interest

There are no conflicts to declare.

## Acknowledgments

This work is supported by a public grant overseen by the French National research Agency as part of the « Investissement d'Avenir » program, through the "IDI 2017" project funded by the IDEX Paris-Saclay, ANR-11-IDEX-0003-02. INRAE, CNRS, UVSQ and Institut Universitaire de France are also gratefully acknowledged for their financial support. DM acknowledges support from Rennes Metropole and Europe (FEDER Fund – CPER PRINT2TAN). Ms. S. Basset, Ms S. Guiheneuf, Dr N. Leclerc and Ms A. Damond are gratefully acknowledged for their help, notably for the ion chromatography, EDX and ESI-MS experiments.

## Notes and references

**A decatungstate-based Ionic liquid exhibiting very low dielectric constant suitable for acting as solvent and catalyst for oxidation of organic substrates**

Yohan Martinetto,[a,b] Bruce Pégot,*[a] Catherine Roch,[a] Mohamed Haouas,[a]
Betty Cottyn-Boitte,[b] Franck Camerel,[c] Jelena Jeftic,[c] Denis Morineau,[d] Emmanuel Magnier,[a] and Sébastien Floquet.*[a]

a. Institut Lavoisier de Versailles, UMR 8180, Université de Versailles St-Quentin en Yvelines, Université Paris-Saclay, 78035 Versailles, France.

b. Institut Jean-Pierre Bourgin, INRAE, Agro Paris Tech, Université Paris Saclay, 78000 Versailles, France.

c. Univ Rennes, CNRS, ISCR (Institut des Sciences Chimiques de Rennes) - UMR 6226, 35000 Rennes, France.

d. Institut de Physique de Rennes, UMR 6251, Université de Rennes 1, 35042 Rennes, France.

Corresponding authors: sebastien.floquet@uvsq.fr, bruce.pegot@uvsq.fr

*Electronic Supporting Information*

# Table of contents:

# Part 1: Experimental section

## 1. General methods

**Fourier Transform Infrared (FT-IR)** spectra were recorded on a 6700 FT-IR Nicolet spectrophotometer, using diamond ATR technique. The spectra were recorded on non-diluted compounds and ATR correction was applied.

**ThermoGravimetric Analysis (TGA)** analysis were recorded on a Seiko TG/DTA 320 thermogravimetric balance. The samples were measured between room temperature and 700 °C (scan rate: 5 °C.min$^{-1}$, under $O_2$).

**Differential scanning calorimetry (DSC)** was performed on a NETZSCH DSC 200 F3 instrument equipped with an $N_2$ cooler, allowing measurements from -170 °C up to 450 °C. The samples were examined at a scanning rate of 10°.min$^{-1}$ by applying two heating and one cooling cycles. The apparatus was calibrated with indium (156.6 °C).

**Ion Chromatography** was performed on a 881 Compact IC pro Metrohm chromatograph using a column Metrosep A Supp 4 - 250/4.0 and a NaHCO3 4mM Na2CO3 1mM water / $CH_3CN$ 75 : 25 mixture as eluant. The flow rate was set at 1mL / min and the temperature at 45°C.

**Electrospray Ionization Mass Spectrometry (ESI-MS)** spectra were collected using a Q-TOF instrument supplied by WATERS. Samples were solubilized in water at a concentration of 10$^{-4}$ M and were introduced into the spectrometer via an ACQUITY UPLC WATERS system whilst a Leucine Enkephalin solution was co-injected via a micro pump as internal standard.

**Energy-dispersive X-ray spectroscopy (EDX)** measurements were performed on a JEOL JSM 5800LV apparatus.

Phase behavior was studied by **Polarized light Optical Microscopy** on a Nikon H600L polarizing microscope equipped with a Linkam "liquid crystal pro system" hotstage LTS420.

**Viscosity measurements** were performed on a Thermofischer Haake MARS III controlled-stress rheometer equipped with a cone-plate geometry (diameter = 35 mm, angle = 1°) and a Peltier thermal regulator.

**Dielectric spectroscopy and conductivity** were performed in the 0.1Hz-1Mhz frequency range. The data are recorded during the cooling of the sample from 70°C to -150°C. The sample was prepared in parallel plate geometry between two brass-plated electrodes with a diameter of 20 mm and a spacing of 60 μm maintained by Teflon spacers. The complex impedance of the as-prepared capacitor was measured with a Novocontrol high resolution dielectric Alpha analyzer with an active sample cell. The temperature of the sample was controlled by a Quatro temperature controller (Novocontrol) with nitrogen as a heating/cooling agent providing a temperature stability better than 0.1°C.

**Nuclear magnetic resonance (NMR)**

$^1$H (300 MHz) NMR and $^{31}$P (121,5 MHz) NMR spectra were recorded at room temperature on a Bruker AC-300 spectrometer in $(CD_3)_2CO$, $CD_3OD$ and $(CD_3)_2SO$. Chemical shifts are reported in parts per million (ppm) relative to internal references. The residual peaks of $(CD_3)_2CO$ (2.05 ppm), $CD_3OD$ (3.31 ppm) or $(CD_3)_2SO$ (2.5 ppm) for $^1$H (300 MHz) NMR spectra.

Liquid $^{183}$W NMR spectra of was obtained on a high resolution 400 MHz Bruker Avance spectrometer, equipped with 10 mm BBO probes. $CD_3CN$ was used as a solvent. Spectra were measured in 10 mm tubes at a Larmor frequency of 16.7 MHz for the $^{183}$W.

## 2. Syntheses and characterizations of POM-ILs

All reagents were purchased from commercial sources and used without further purification.

**Synthesis of TBA$_4$[W$_{10}$O$_{32}$].** TBA$_4$[W$_{10}$O$_{32}$] was synthetised as described in the literature.[1] **IR/cm$^{-1}$** (see Figure S1): 2955 (s), 2928 (vs), 2858 (s), 1469 (m), 1380 (w), 1162 (vw), 994 (vw), 958 (s), 891 (s), 805 (vs), 586 (w), 435 (m), 404 (m), 335 (w). **$^{183}$W NMR** (See Figure S5, Supporting Information) (16.7 MHz, CD$_3$CN): δ (ppm) -22.9 (s, 8W), -166 (s, 2W).

**Synthesis of POM-IL (P$_{6,6,6,14}$)$_4$[W$_{10}$O$_{32}$]:** This synthesis was inspired from Fournier[1] In a 250 mL beaker flask sodium tungstate dihydrate (16 g, 50 mmol) was dissolved in 100 mL of boiling distilled water. Then 33.5 mL of boiling HCl (3 M) was added with a rapid stirring. After 2 minutes of strong boiling, tetradecyltrihexylphosphonium chloride (P$_{6,6,6,14}$,Cl) (7.55 g, 15 mmol) in 10 mL of ethanol was added. Polyoxometalate-based Ionic Liquid formed a dense phase in the bottom of the beaker. Finally, the aqueous phase was separated, and the POM-IL phase was washed 3 times with 40 mL of boiling distilled water and dried with a vacuum pump until the POM-IL became a clear blue-green viscous liquid. Yield 15 g, 72% based on tungstate. **IR/cm$^{-1}$** (see Figure S1): 2954 (s), 2926 (vs), 2855 (s), 1466 (m), 1408 (w), 1378 (w), 1212 (vw), 1112 (vw), 723 (m), 994 (vw), 958 (s), 891 (s), 805 (vs), 586 (w), 435 (m), 404 (m), 348 (vw), 335 (w). **$^1$H NMR** (See Figure S4) (300 MHz, (CD$_3$)$_2$CO): δ (ppm) 2.48 (m, 8H), 1.71 (m, 8H), 1.56 (m, 8H), 1.4-1.2 (m, 34H) and 0.88 (m, 12H). **$^{31}$P NMR** (121,5 MHz, (CD$_3$)$_2$CO): δ (ppm) 33,95. **$^{183}$W NMR** (See Figure S5) (16.7 MHz, (CD$_3$CN)): δ (ppm) -20.9 (s, 8W), -163 (s, 2W). **TGA** (see Figure S): A weight loss of 39.4% between RT and 700 °C corresponds to a combustion of 4 cations (P$_{6,6,6,14}$)$^+$ (expected 39.2 %) Thermogravimetric analyses show that the compound is stable up to 200 °C without any degradation and a total absence of water. **EDX** : no trace of Na and Cl coming from the tungstate or the phoshonium precursors detected. By **ion chromatography**, no trace of chloride ions has been detected. **ESI-MS** (CH$_3$CN, see Figure S2) : m/z 587.4 ([W$_{10}$O$_{32}$]$^{4-}$ expected m/z 587.6) and 783.8([HW$_{10}$O$_{32}$]$^{3-}$ expected m/z 783.8). **DSC**: The clear glass transition was centered around Tg = -10 °C.

Note that for this compound and other POM-ILs, an excess of (P$_{6,6,6,14}$)Cl can sometimes be identified by TGA, EDX and Ion chromatography experiments. TGA experiment is senssitive enough to get this information quickly. If so, the compound is dissolved in THF and passed through a steric exclusion gel column. The bigger species, *i.e.* the expected POM-IL, is not retained by the column and passes first. The ($P_{6,6,6,14}$)Cl in excess are smaller and are therefore retained by the gel. This technique allows to easily purify our POM-ILs. The purified compound is then dried under vacuum before characterization by FT-IR, TGA, Ion chromatography, EDX and so on.

**Synthesis of ($P_{6,6,6,14}$)$_2$[$W_6O_{19}$]:** This synthesis was performed as described by Fournier.[2] A mixture of 16,5 g (50 mmol) of sodium tungstate dihydrate, 20 mL of acetic anhydride, and 15 mL of DMF is stirred in a 125 mL Erlenmeyer flask at 100 °C for 3 h. A solution of 10 mL of acetic anhydride and 9 mL of 12N HCl in 25 mL of DMF is added with stirring and the mixture is gravity filtered through a medium porosity filter paper in order to eliminate the undissolved white solid. After washing the solide with 25 mL of methanol, the clear filtrate is allowed to cool to room temperature. A solution of tetradecyltrihexylphosphonium chloride ($P_{6,6,6,14}$,Cl) (9.1 g, 18 mmol, 2.1 eq.) in 10 mL of ethanol was added. Polyoxometalate-based Ionic Liquid formed a dense phase in the bottom of the beaker. Finally, the organic phase was separated, and the POM-IL phase was dried with a vacuum pump until the POM-IL became a clear blue-green viscous liquid. Yield 1.5 g, 8% based on tungstate. **IR/cm$^{-1}$** (see Figure S1): 2955 (s), 2927 (vs), 2855 (s), 1465 (m), 1409 (w), 1378 (w), 1214 (vw), 1111 (vw), 723 (m), 978 (vs), 891 (vw), 815 (vs), 720 (w), 586 (m), 446 (s), 405 (vw), 368 (w). **$^{183}$W NMR** (See Figure S5) (16.7 MHz, (CD$_3$CN)): δ (ppm) 56 (s, 6W).

# Part 2: Characterizations by FT-IR spectra

FT-IR spectra comparison of **(P$_{6,6,6,14}$)Cl, TBA$_4$[W$_{10}$O$_{32}$], (P$_{6,6,6,14}$)$_4$[W$_{10}$O$_{32}$]** and **(P$_{6,6,6,14}$)$_2$[W$_6$O$_{19}$]** are given in Figure S1.

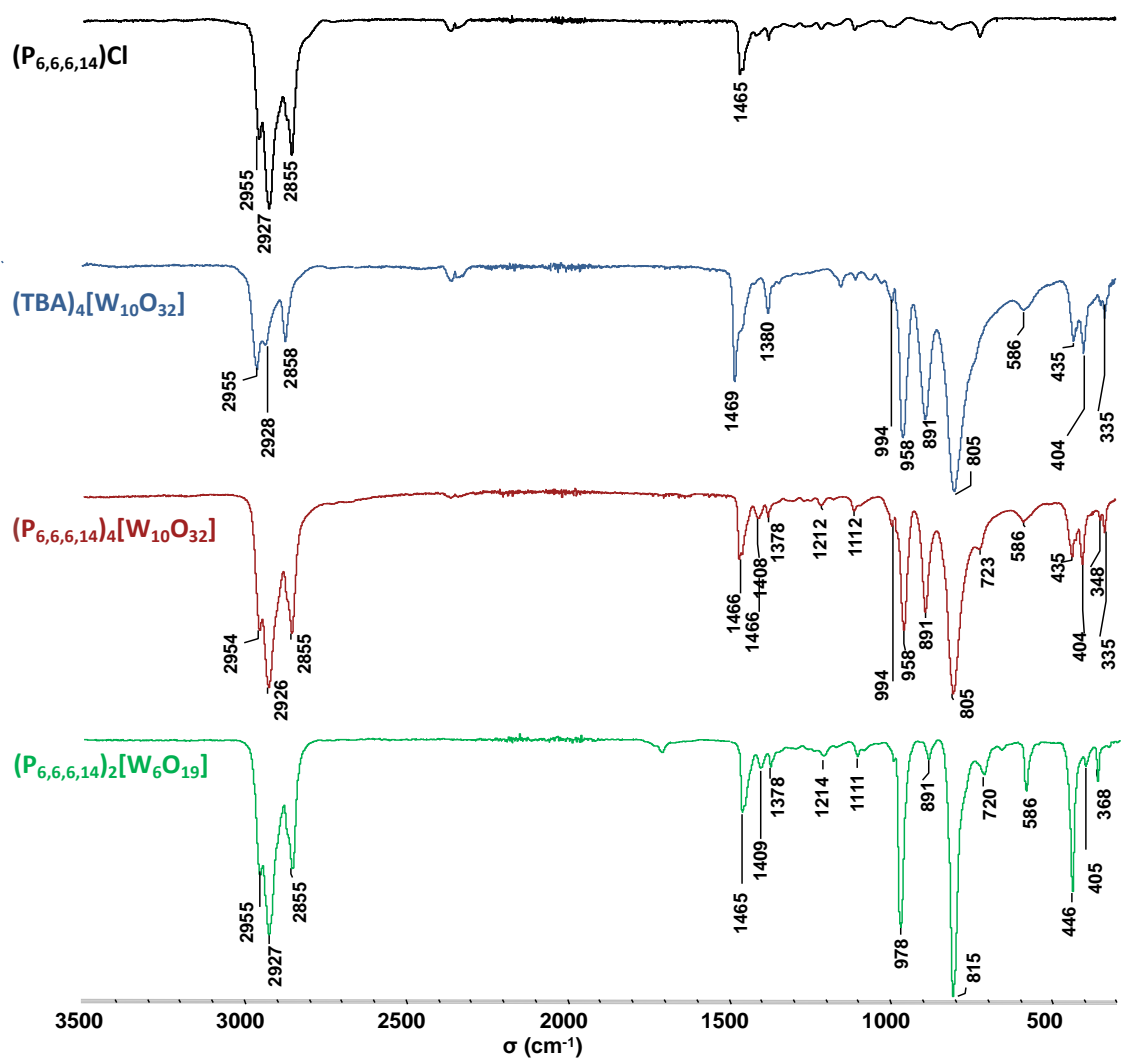

*Figure S1: Infrared spectra comparing (P$_{6,6,6,14}$)$_4$[W$_{10}$O$_{32}$], (P$_{6,6,6,14}$)$_2$[W$_6$O$_{19}$], (TBA)$_4$[W$_{10}$O$_{32}$] as reference and (P$_{6,6,6,14}$)Cl.*

All of the vibration bands between 3000 and 2850 cm$^{-1}$ were assigned at the C-H aliphatic bonds.

- **(P$_{6,6,6,14}$)$_4$[W$_{10}$O$_{32}$]**

In the FT-IR spectrum, the bands attributed to the [W$_{10}$O$_{32}$]$^{4-}$ are respectively : 958 cm$^{-1}$ for W-O$_t$ vibration and 891, 805 and 586 cm$^{-1}$ for W-O$_b$-W and W-O$_c$-W vibration. These vibration bands are in agreement with literature[1] and the TBA$_4$[W$_{10}$O$_{32}$] reference spectra.

- **(P$_{6,6,6,14}$)$_2$[W$_6$O$_{19}$]**

The bands assigned to the [W$_6$O$_{19}$]$^{2-}$ are respectively : 978 cm$^{-1}$ for W=O vibration and 815, and 446 cm$^{-1}$ for W-O vibration. These vibration bands are in agreement with literature.[2]

# Part 3: Characterizations by ESI-MS

The ESI-MS spectrum of compound $(P_{6,6,6,14})_4[W_{10}O_{32}]$ was recorded in $CH_3CN$ in negative mode. As shown in Figure S2, the spectrum exibits two major peaks at m/z 587.4 and 783.8; which are attributed to the species $[W_{10}O_{32}]^{4-}$ (587.6) and $[HW_{10}O_{32}]^{3-}$ (783.8), respectively. It confirms the nature of the POM in the compound $(P_{6,6,6,14})_4[W_{10}O_{32}]$. Note that the peak found at m/z 703.8 is attributed to the formation of $[W_6O_{19}]^{2-}$ generated during the ESI-MS experiment. This assumption is confirmed by the fact that by increasing the cone voltage of the experiment this peak becomes predominant.

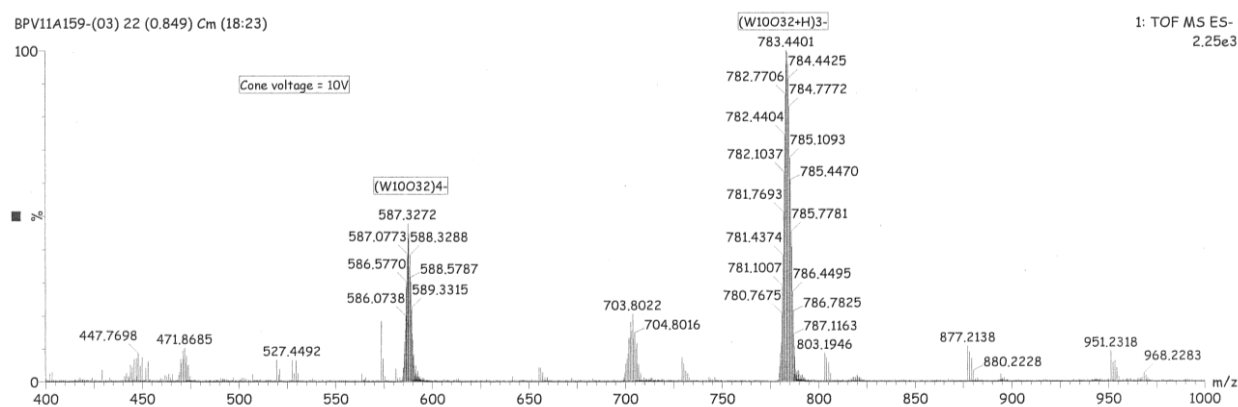

*Figure S2: ESI-MS spectrum of $(P_{6,6,6,14})_4[W_{10}O_{32}]$ in $CH_3CN$ in negative mode.*

# Part 4: ThermoGravimetric Analysis (TGA)

The TGA experiment is fully informative for the characterization of POM-ILs. Considering that under $O_2$ the compounds are burned into $WO_3$ and $P_2O_5$ above 600°C and that all the organic parts and eventually chloride ions coming from an excess of $(P_{6,6,6,14})Cl$ are removed during the TGA experiments.

As examplified below, we can easily distinguish samples with an excess of $(P_{6,6,6,14})Cl$. In Figure S3a, the TGA curve recorded for a POM-IL indicates a weight loss of 47%, which fully agree with the formula $(P_{6,6,6,14})_4[W_{10}O_{32}].((P_{6,6,6,14})Cl)_{1.6}$. EDX measurement confirms the presence of chloride in excess in the same order. In this case, the compound can be purified by passing THF solution of the compound on steric exclusion gel column until the TGA agrees with the expected weight loss (Figure S3b, which exactly corresponds to the weight loss expected for $(P_{6,6,6,14})_4[W_{10}O_{32}]$.(calculated 39.2%). The purity, and the absence of supplementary $(P_{6,6,6,14})Cl$ salt is confirmed by EDX and by ion chromatography.

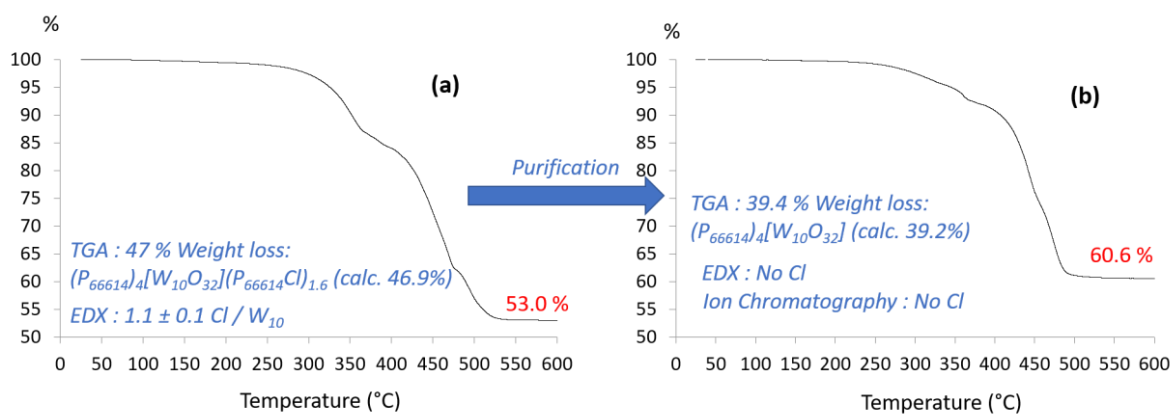

*Figure S3: ThermoGravimetric curve of $(P_{6,6,6,14})_4[W_{10}O_{32}]$ recorded under oxygen at 5 °C/minute.*

The TGA curve under oxygen flow confirms the absence of water solvate in the POM-IL due to the missing mass drop around 100 °C. We can also propose an approximate value of decomposition around 200 °C.

A final weight of 60.6% perfectly matches with the combustion of the organic part of the 4 organic cations $(P_{6,6,6,14})^+$ for yielding at the end $P_2O_5$ and $WO_3$.

## Part 5: NMR studies in solution

1- $^1$H NMR: Characterization of $(P_{6,6,6,14})_4[W_{10}O_{32}]$ comparing to $(P_{6,6,6,14})Cl$ in $(CD_3)_2CO$.

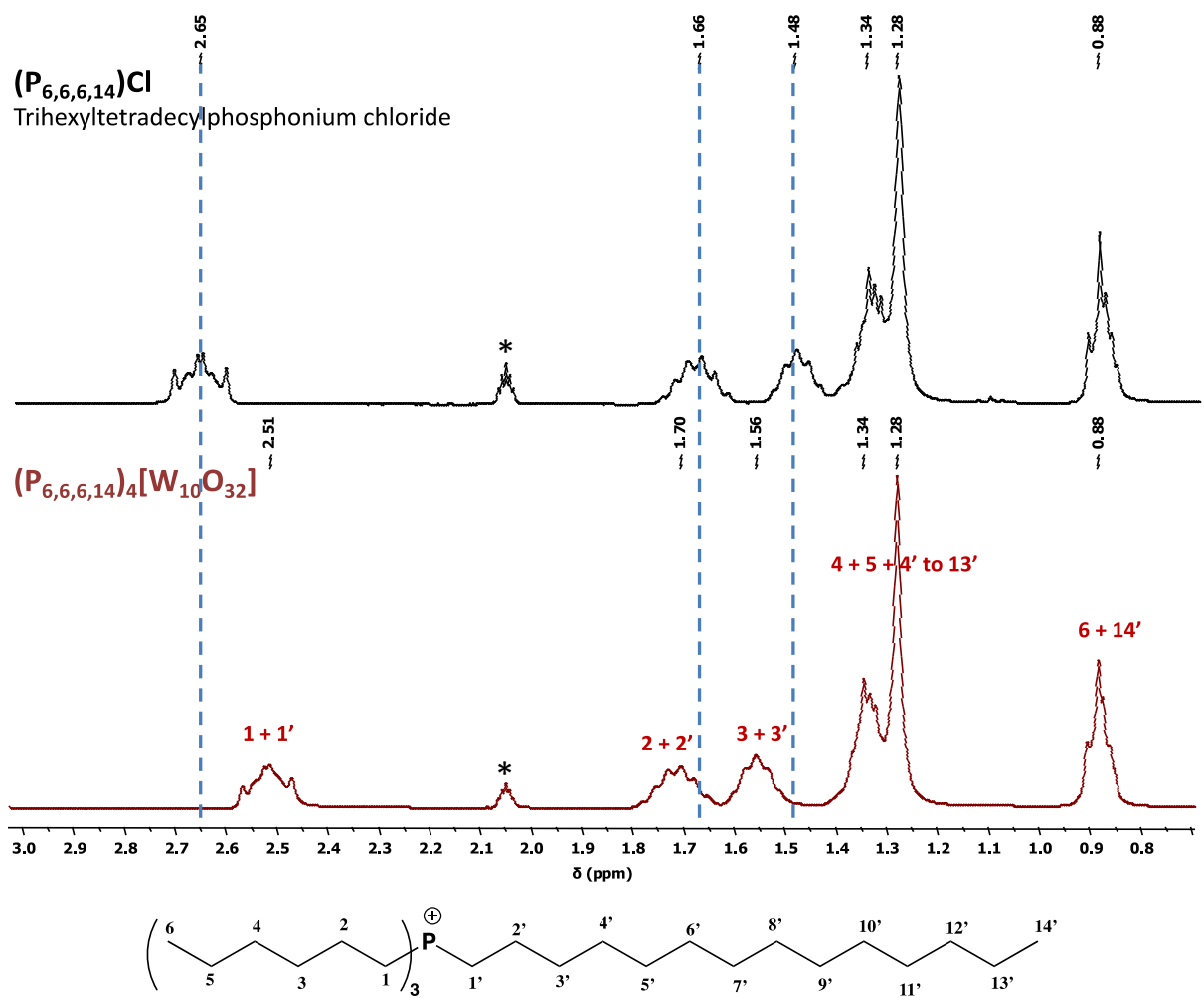

Figure S4: $^1$H NMR spectra comparing $(P_{6,6,6,14})_4[W_{10}O_{32}]$ and $(P_{6,6,6,14})Cl$ and peaks' attibution. * indicate the signal attributed to $(CD_3)_2CO$

The $^1$H NMR spectra indicate significant shifts in both directions of the methylenic protons (2.51 ppm, 1.70 ppm and 1.56 ppm) which are closest from the phosphorus atom. This gap suggests a strong interaction between the POM and cations.

2- **$^{183}$W NMR (16.7 MHz, CD$_3$CN):** NMR spectra comparing (P$_{6,6,6,14}$)$_4$[W$_{10}$O$_{32}$], (P$_{6,6,6,14}$)$_2$[W$_6$O$_{19}$] and (TBA)$_4$[W$_{10}$O$_{32}$].

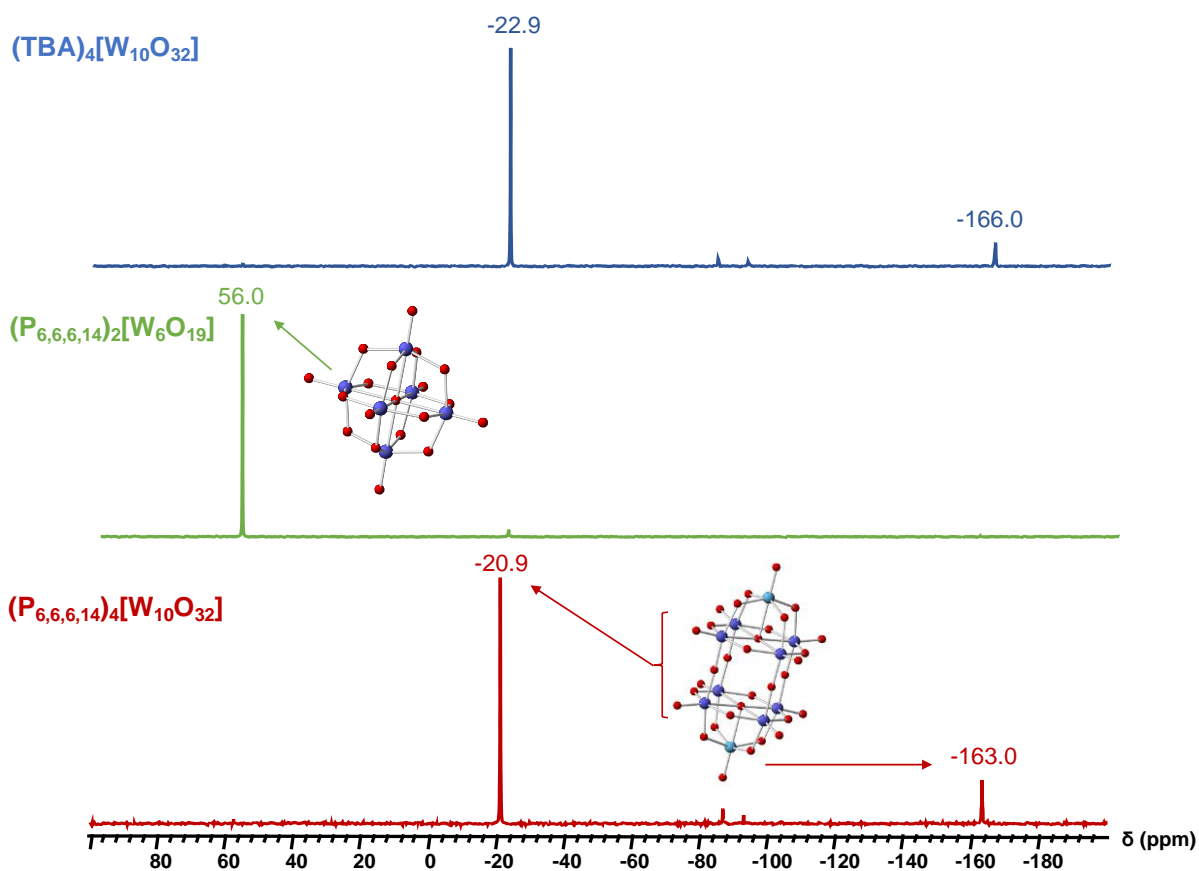

*Figure S5: $^{183}$W NMR (16.7 MHz, CD$_3$CN) spectra comparing (P$_{6,6,6,14}$)$_4$[W$_{10}$O$_{32}$], (P$_{6,6,6,14}$)$_2$[W$_6$O$_{19}$] and (TBA)$_4$[W$_{10}$O$_{32}$].*

The $^{183}$W NMR spectrum of **(P$_{6,6,6,14}$)$_4$[W$_{10}$O$_{32}$]** shows two peaks around -21 and -163 ppm with a 4: 1 intensity ratio, which correspond to two expected types of tungsten atoms, equatorial and capped, in the [W$_{10}$O$_{32}$]$^{4-}$ structure reported in the literature as TBA salt.[3] For [W$_6$O$_{19}$]$^{2-}$, the highly symmetrical Lindqvist anion produce a single pic at 56 ppm in $^{183}$W NMR and confirms its formation.[4] No trace of [W$_6$O$_{19}$]$^{2-}$ was observed in the spectrum of (P$_{6,6,6,14}$)$_4$[W$_{10}$O$_{32}$].

# Part 6: Rheological analyses

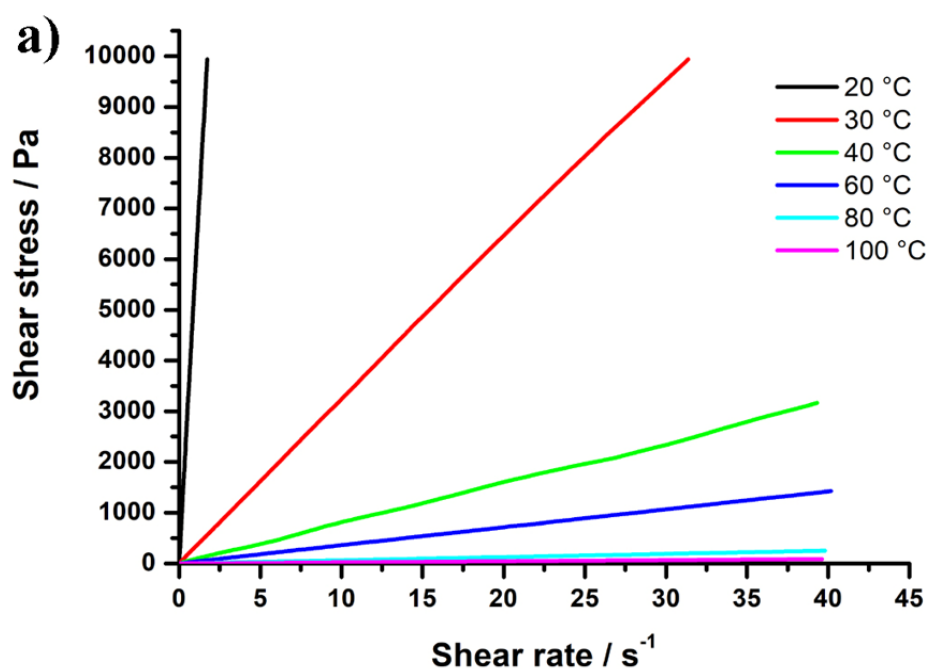

*Figure S6: Analysis of $(P_{6,6,6,14})_4[W_{10}O_{32}]$. a) Flow curves $\tau = f(\dot{\gamma})$ showing the evolutions of the shear stress versus the shear rate at various temperatures for this compound. Each line corresponds to a curve $\tau = f(\dot{\gamma})$ for a given temperature;*

### Conductivity of $(P_{6,6,6,14})_4[W_{10}O_{32}]$.

The dielectric loss as a function of the frequency for a wide range of studied temperatures are shown in Figures S7 and S8. The temperature step between each curve is 5 °C. For the highest temperature T = 70 °C (upper curve), the intensity of the dielectric loss is dominated by ionic conductivity. The signal flattens at lower temperature and the dipolar relaxation modes show up on cooling down to the lowest studied temperature -150 °C (lower curve), as also shown in Figure S8.

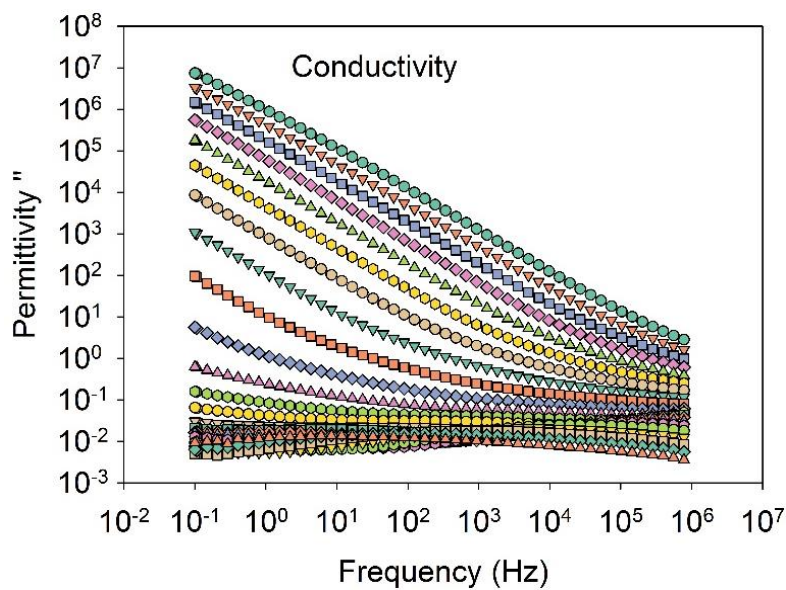

*Figure S7: Dielectric loss as a function of the frequency at regularly spaced temperatures ranging from T = 70 °C (upper curve) to -150 °C (lower curve).*

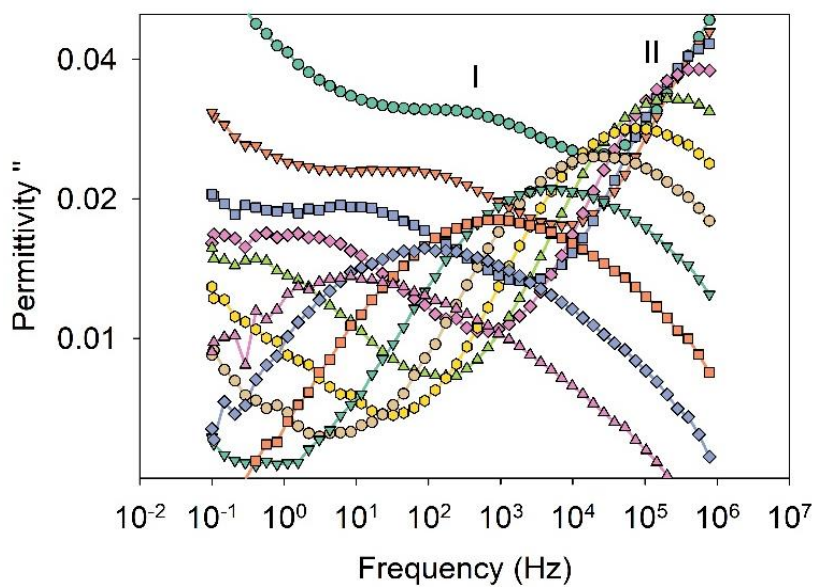

*Figure S8: Magnified plot of the dielectric loss as a function of the frequency in the region of low temperatures from T = -50 °C (upper curve) to -150 °C (lower curve).*

The ionic conductivity may be related to the diffusion coefficient through the Nernst-Einstein equation $\sigma_0 = \frac{ne^2 D}{kT}$, where *n* is the number density of charge carriers, *e* is the electric charge and *k* is Boltzmann's constant. In addition, the diffusion coefficient may be related to the viscosity $\eta$ through the Stoke-Einstein equation $D = \frac{kT}{6\pi a \eta}$, where *a* is the hydrodynamic radius of the diffusion particles. If both equations apply, the conductivity is in turn inversely proportional to the viscosity. This relation between conductivity and viscosity was actually confirmed, as shown in Figure S9, where both quantities are plotted as a function of the inverse temperature. In the temperature range above room temperature where both quantities were measured, they fulfill an Arrhenius law with a comparable value of the activation energy (85 ± 5 kJ.mole$^{-1}$). At lower temperature, one noticed a deviation of the inverse conductivity, which exhibits a super-Arrhenius behavior. This apparent increase of the activation energy is typically observed in glassforming liquids and commonly attributed to the onset of dynamical cooperativity in fragile liquids at low temperature under supercooling conditions.[5]

The two relaxation modes exhibit significantly smaller activation energies (57 and 32 kJ.mole$^{-1}$ for the mode I and II respectively), and a Arrhenius-like behaviour. This suggest that, unlike conductivity, these modes are decoupled from the hydrodynamic viscosity.[6] They are more likely attributed to local rearrangements, usually named beta-processes, that induce polarization fluctuations, such as ligands tumbling or POM–ligand breathing modes. This hypothesis is comforted by the small values of the dielectric strength of the two modes ($0.1 < \Delta\varepsilon_{1,2} < 0.2$ ), in agreement with the overall spherical symmetry of the POM and very weakly polar character of the ligands.

It is commonly observed for glassforming liquids that the extrapolation of the main structural relaxation time reaches values of the order of $10^2$ s at the calorimetric glass transition temperature Tg.[7,8] Universal scaling of conductivity, viscosity and the main structural relaxation were also observed for series of ionic liquids.[7] Although the present liquid also demonstrates a coupling between conductivity and viscosity, it is noteworthy that the two dielectrically active relaxation modes extrapolate to $10^2$ s for tempartures that are far below the calorimetric Tg. This further confirms that they correspond to secondary (beta) processes. The experimental detection of an electrically active mode associated with the main (alpha) relaxation was presumably hidden due to the strong contribution of superimposed conductivity.

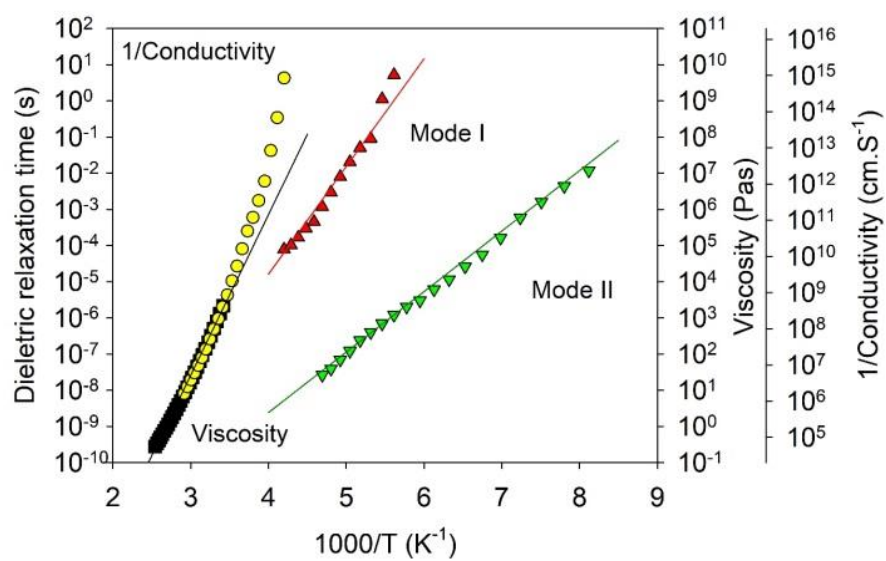

*Figure S9: Arrhenius plot of the inverse conductivity (circles), dielectric relaxation time of the mode I (upward triangles) and mode II (downward triangles), and viscosity (squares). Solid lines are Arrhenius fits to the data.*

# Part 7: Catalysis experiments

1- **Experimental procedure for oxidation of alcohols.**

The experimental procedure used for catalysis is depicted in scheme S1 below, while Table S1 give some optimization reaction leading to the protocol wh chose to use for this study.

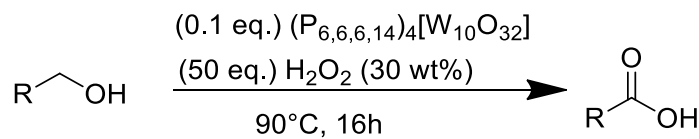

*Scheme S1: General equation reaction of oxidation of alcohols*

*Table S1: optimisation and test reactions.*

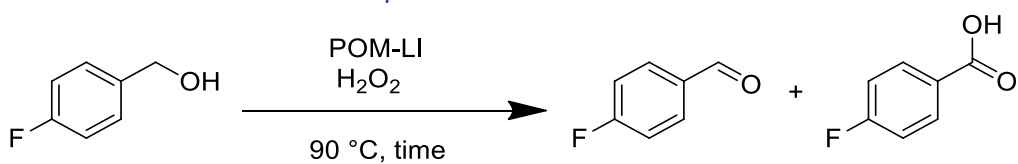

| Entry[a] | Catalyst $(P_{6,6,6,14})_4[W_{10}O_{32}]$ | $H_2O_2$ 30% W | Time (h) | Conversion (%)[a] | Aldehyde yield (%)[a] | Acid yield (%)[a] |
|---|---|---|---|---|---|---|
| 1[b] | 0.1 eq. | 50 eq. | 1 | 88 | 24 | 56 |
| 2 | 0.1 eq. | 50 eq. | 16 | 100 | / | 98 (93)[b] |
| 3[c] | 0.1 eq. | 0 eq. | 16 | 4 | <1 | <1 |
| 4 | 0 eq. | 50 eq. | 16 | 15 | <1 | 10 |

[a]determined by $^{19}$F NMR analysis using trifluoroanisol as internal standard. [b]Isolated yield after product separation by a reusable steric polymer column.

**Protocol followed for this study :** In a 100 mL flask equipped with condenser 1 eq. of alcohol and 0.1 eq. of catalyst $((P_{6,6,6,14})_4[W_{10}O_{32}])$ were introduced. The mixture was stirred during a few minutes at 90 °C and then 50 eq. of hydrogen peroxide (30 wt% in water) were quickly added. This biphasic system was kept at 90°C for 16 h. The biphasic mixtures were then cooled at room temperature and separated by decantation after addition of 20 mL of distilled water. The aqueous phase was extracted 3 times with 20 mL of diethyl ether and the organic phase obtained was dried over MgSO4 and concentrated under reduced pressure. Considering the ionic liquid phase, acetone was added until a homogenous phase was obtained (around 20 mL). The solution was dried over MgSO4 and concentrated under reduced pressure. These two collected organic phases were mixed, and a minimum of tetrahydrofuran was added until a clear solution was obtained. The catalyst was then separated from the reactions products by a reusable steric exclusion polymer column composed of poly(styrene-co-divinylbenzene) eluted by tetrahydrofuran (around 100 mL). the POM-IL, was firstly recovered and after evaporation of THF the catalyst could be directly reused in an other cycle. The oxidation products are isolated secondly from the column. Purification of reaction products, when necessary, was performed by recrystallization, pentane washing or silica plate. All data of the synthetized product are in accordance with aldrich commercially available componds and their analytical documents (see Aldrich website).

2- Reference reactions

To evidence the catalytic activity of $(P_{6,6,6,14})_4[W_{10}O_{32}]$, number of reference reactions were performed. The experimental conditions are given in Tables S2 and S3.

*Table S2: Table of verification reactions, general protocol with different catalysts and conditions.*

| Entry | Catalyst | Aldehyde yield (%) | Acid Yield (%) |
|---|---|---|---|
| 1 | No catalyst | <1 | 10[c] |
| 2 | $(P_{6,6,6,14})_4Cl$ | 7 | 23[c] |
| 3[b] | $(P_{6,6,6,14})_4[W_{10}O_{32}]$ | <1 | <1[c] |
| 4 | $(P_{6,6,6,14})_4[W_{10}O_{32}]$ | / | 93[d] |

[a]Reaction conditions: 4-fluorobenzyl alcohol (1 eq.), H$_2$O$_2$ 30 wt% (50 eq.), catalyst (0.1 eq.), 90 °C, overnight.
[b]Reaction conditions: 4-fluorobenzyl alcohol (1 eq.), without H$_2$O$_2$, catalyst (0.1 eq.), 90 °C, overnight.
[c]None isolated yield, detected by $^{19}$F NMR analysis.
[d]Isoleted yield after catalyst/product separation by a reusable steric polymer column and purification by pentane washing.

*Table S3: Reactivity comparison between $(P_{6,6,6,14})_4[W_{10}O_{32}]$ and $(P_{6,6,6,14})_2[W_6O_{19}]$.*

| Entry | Catalyst | Alcohol conversion (%)[b] | Unknow product (%)[b] | Aldehyde (%)[b] | Acid (%)[b] |
|---|---|---|---|---|---|
| 1 | $(P_{6,6,6,14})_2[W_6O_{19}]$ | 77 | 20 | 24 | 26 |
| 2 | $(P_{6,6,6,14})_4[W_{10}O_{32}]$ | 100 | 19 | 24 | 56 |

[a]Reaction conditions: 4-fluorobenzyl alcohol (1 eq.), H$_2$O$_2$ 30 wt% (50 eq.), catalyst (0.1 eq.), 90 °C, 2 h.
[b]None isolated yield, only detected by $^{19}$F NMR analysis.

The Table S2 collects data from different verifications reactions tested. We can see in Entry 1 a reaction without catalyst, Entry 2 a reaction without $[W_{10}O_{32}]^{4-}$, Entry 3 a reaction without H$_2$O$_2$ and in Entry 4 the "classical" reaction in POM-IL medium in presence of H$_2$O$_2$. This table demonstrates that the couple $(P_{6,6,6,14})_4[W_{10}O_{32}]$ / H2O2 is needed to observe catalysis with a good yield.

Besides, the Table S3, demonstrates that of $(P_{6,6,6,14})_2[W_6O_{19}]$ is less efficient than $(P_{6,6,6,14})_4[W_{10}O_{32}]$, which explains the decrease of the yield after several runs, concomitantly with the formation of $(P_{6,6,6,14})_2[W_6O_{19}]$ (vide infra).

### 3- Recyclability tests of catalyst and analysis after catalysis

The integrity of the catalyst after catalytic process was checked by $^{183}$W NMR (Figure S9). The $^{183}$W NMR spectrum after 5 runs shows the $(P_{6,6,6,14})_4[W_{10}O_{32}]$ POM is maintained while a minor quantity of $(P_{6,6,6,14})_2[W_6O_{19}]$ linqvist POM (around 21 % after 5 runs) appears, in agreement with the decrease of the catalytic yield (*vide supra*).

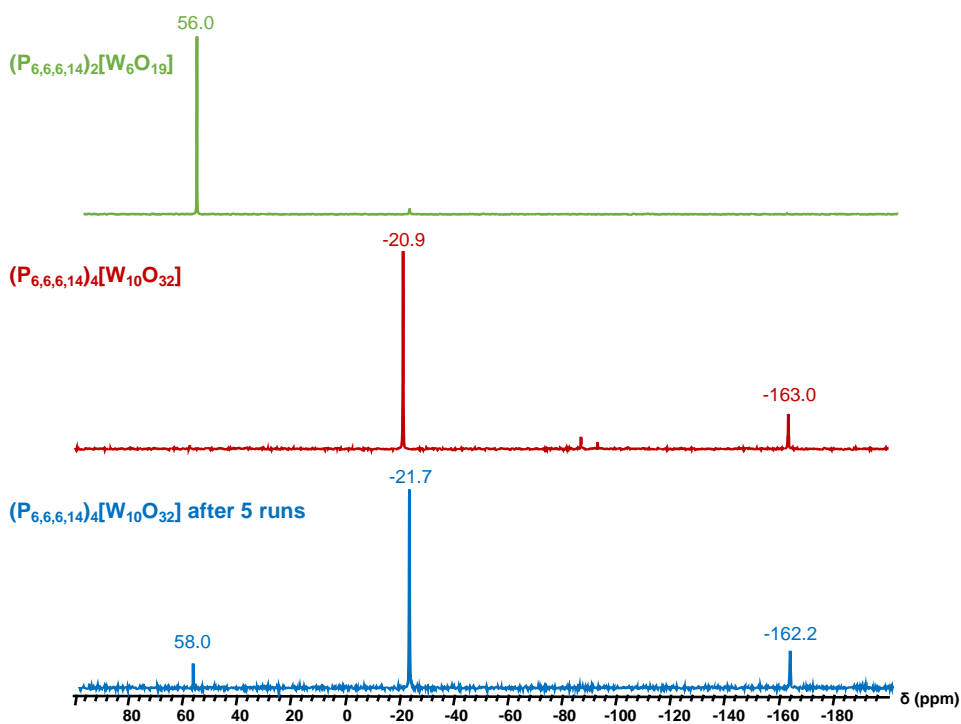

Figure S10: $^{183}$W NMR (16.7 MHz, CD$_3$CN) spectra comparing $(P_{6,6,6,14})_4[W_{10}O_{32}]$ after 5 runs of catalysis, initial $(P_{6,6,6,14})_4[W_{10}O_{32}]$ and $(P_{6,6,6,14})_2[W_6O_{19}]$.